\documentclass[11pt]{llncs}

\usepackage{graphicx, amsmath, amssymb, cite}
\newtheorem{problemEnv}[theorem]{Problem}{\bfseries}{\itshape}

\newcommand{\modularity}[1][\clusteringSymb]{\operatorname{Q}(#1)}

\newcommand{\clusteringSymb}{\mathcal{C}}

\newcommand{\modul}{{\sc Modularity}}

\newcommand{\parti}{{\sc 3-Partition}}

\begin{document}

\title{Maximizing Modularity is hard\thanks{This work was partially
    supported by the DFG under grants BR~2158/2-3, WA~654/14-3,
    Research Training Group~1042 ``Explorative Analysis and
    Visualization of Large Information Spaces'' and by EU under grant
    DELIS (contract No.~001907).}}
\author{%
  Ulrik Brandes\inst{1} \and%
  Daniel Delling\inst{2} \and%
  Marco Gaertler\inst{2} \and%
  Robert G\"orke\inst{2} \and%
  Martin Hoefer\inst{1}%
  \thanks{Direct all correspondence to {\tt
      hoefer@inf.uni-konstanz.de}} \and%
  Zoran Nikoloski\inst{3} \and%
  Dorothea Wagner\inst{2}}

\institute{%
  Department of Computer \& Information Science, University of
  Konstanz, Germany 
  \and
  Faculty of Informatics, Universit{\"a}t Karlsruhe (TH), Germany
  \and
  Department of Applied Mathematics, Faculty of Mathematics and
  Physics,\\ Charles University, Prague, Czech Republic
}

\maketitle

\begin{abstract}
Several algorithms have been proposed to compute partitions of networks
into communities that score high on a graph clustering index called
modularity.  While publications on these algorithms typically contain
experimental evaluations to emphasize the plausibility of results, none of
these algorithms has been shown to actually compute optimal partitions.
We here settle the unknown complexity status of modularity maximization
by showing that the corresponding decision version is NP-complete in the
strong sense.  As a consequence, any efficient, i.e.\ polynomial-time,
algorithm is only heuristic and yields suboptimal partitions on many
instances.
\end{abstract}
%
\section{Introduction}

Partioning networks into communities is a fashionable statement of 
the graph clustering problem, which has been studied for decades
and whose applications abound.

Recently, a new graph clustering index called \emph{modularity}
has been proposed~\cite{ng-fecsn-04}.  It immediately
prompted a number of follow-up studies concerning different
applications and possible adjustments of the measure (see, e.g.,
\cite{fb-rlcd-06,zmw-itanm-05,mrc-lmmnc-05,fpp-spcn-06}).  Also, a wide
range of algorithmic approaches approaches has been considered, for
example based on a greedy agglomeration~\cite{n-fadcs-03,cnm-fcsln-04},
spectral division~\cite{n-mcsn-06,ws-scacg-03}, simulated
annealing~\cite{gsa-mfrg-04,rb-smcd-06} and extremal
optimization~\cite{da-cdeo-05}.

None of these algorithms, however, has been shown to be produce optimal
partitions.  While the complexity status of modularity maximization is
open, it has been speculated~\cite{n-mcsn-06} that it might be NP-hard
due to similarity with the MAX-CUT problem.

In this paper, we provide the first complexity-theoretic argument as
to why the problem of maximizing modularity is intractable by proving
that it is NP-complete in the strong sense.  This means that there is no
correct polynomial-time algorithm to solve this problem for every instance
unless P~=~NP.   Therefore, all of the above algorithms eventually deliver
suboptimal solutions, and there is no hope for an efficient algorithm
that computes maximum modularity partitions on all problem instances.
In a sense, our result thus justifies the use of heuristics for modularity
optimization.

\section{Modularity}

Modularity is a quality index for clusterings defined as follows. We
are given a simple graph $G = (V,E)$, where $V$ is the set of vertices
and $E$ the set of (undirected) edges. If not stated otherwise, $n =
|V|$ and $m = |E|$ throughout. The \emph{degree} $\deg(v)$ of a vertex
$v \in V$ is the number of edges incident to $v$. A \emph{cluster} or
community $C \subseteq V$ is a subset of the vertices. A \emph{clustering}
$\clusteringSymb = \{C_1,\ldots,C_t\}$ of $G$ is a partition of $V$ into
clusters such that each vertex appears in exactly one cluster. With
a slight disambiguation, the \emph{modularity}~\cite{ng-fecsn-04}
$\modularity[\clusteringSymb]$ of a clustering $\clusteringSymb$ is
defined as
\begin{equation}\label{eq:modularity}
 \modularity[\clusteringSymb] = \sum_{C \in \clusteringSymb} \left[
\frac{|E(C)|}{m} - \left(\frac{|E(C)| + \sum_{C' \in
    \clusteringSymb} |E(C,C')|}{2m} \right)^2\right]~,
\end{equation}
where $E(C,C')$ denotes the set of edges between vertices in clusters $C$
and $C'$, and $E(C)=E(C,C)$. Note that $C'$ ranges over all clusters, so
that edges in $E(C)$ are counted twice in the squared expression. This
is to adjust proportions, since edges in $E(C,C')$, $C \neq C'$, are
counted twice as well, once for each order of the arguments.  Note that
we can rewrite Eq.~\eqref{eq:modularity} into the more convenient form
\begin{equation}\label{eq:modularity2}
\modularity[\clusteringSymb] =
\sum_{C \in \clusteringSymb} \left[
\frac{|E(C)|}{m} - \left(\frac{\sum_{v \in C}
  \deg(v)}{2m}\right)^2\right]~.
\end{equation}
It reveals an inherent trade-off: to maximize the first term, many
edges should be contained in clusters, whereas minimization of the
second term is achieved by splitting the graph into many clusters of
small total degrees.  In the remainder of this paper, we will make use
of this formulation.

\section{NP-Completeness}

To formulate our complexity-theoretic result, we need to consider 
the following decision problem underlying modularity maximization.

\begin{problemEnv} [\modul]
Given a graph $G$ and a number $K$, is there a clustering
$\clusteringSymb$ of $G$, for which $\modularity[\clusteringSymb] \ge K$?
\end{problemEnv}
%
Note that we may ignore the fact that, in principle,
$K$ could be a real number in the range $[0,1]$, because
$4m^2\cdot\modularity[\clusteringSymb]$ is integer for every partition
$\clusteringSymb$ of $G$ and polynomially bounded in the size of $G$.

Note also that modularity maximization cannot be easier than the decision
problem, because determining the maximum possible modularity index of
a graph immediately yields an answer to the decision question.

Our hardness result for \modul\ is based on a transformation from the
following decision problem.

\begin{problemEnv} [\parti]
Given $3k$ positive integer numbers $a_1,\ldots,a_{3k}$ such that the
sum $\sum_{i=1}^{3k} a_i = kb$ and $b/4 < a_i < b/2$ for an integer
$b$ and for all $i=1,\ldots,3k$, is there a partition of these numbers
into $k$ sets, such that the numbers in each set sum up to $b$?
\end{problemEnv}
%
We will show that an instance $A=\{a_1,\ldots,a_{3k}\}$ of \parti\
can be transformed into an instance $(G(A),K(A))$ of \modul, such
that $G(A)$ has a clustering with modularity at least $K(A)$, if and
only if $a_1,\ldots,a_{3k}$ can be partitioned into $k$ sets of sum
$b = \frac{1}{k}\sum_{i=1}^k a_i$ each. 

It is crucial that \parti\ is \emph{strongly}
NP-complete~\cite{gj-crmsu-75}, i.e.\ the problem remains NP-complete
even if the input is represented in unary coding.  This implies that no
algorithm can decide the problem in time polynomial even in the sum of
the input values, unless $P=NP$.  More importantly, it implies that our
transformation need only be pseudo-polynomial.

\begin{figure}
\centering
\includegraphics[width=.6\linewidth]{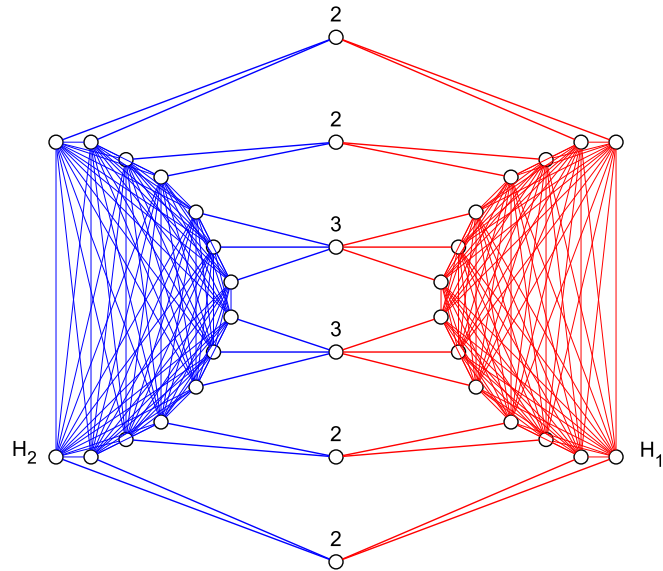}
\caption{\label{fig:NPC:construct} An example graph $G(A)$ for the
  instance $A = \{2,2,2,2,3,3\}$ of \parti. Edge colors indicate
  edges to and within the $k=2$ cliques $H_1$ (red) and $H_2$
  (blue). Vertex labels indicate the corresponding numbers $a_i \in
  A$.}
\end{figure}

The reduction is defined as follows.  From an instance~$A$ of \parti,
construct a graph $G(A)$ with $k$ cliques (completly connected
subgraphs) $H_1,\ldots,H_k$ of size $a = \sum_{i=1}^{3k} a_i$
each. For each element $a_i \in A$ we introduce a single \emph{element
  vertex}, and connect it to $a_i$ vertices in each of the $k$ cliques
in such a way that each clique member is connected to exactly one
element vertex. It is easy to see that each clique vertex then has
degree $a$ and the element vertex corresponding to element $a_i\in A$
has degree $ka_i$. The number of edges in $G(A)$ is $m =
\frac{k}{2}a(a+1)$. See Fig.~\ref{fig:NPC:construct} for an example.
Note that the size of $G(A)$ is polynomial in the unary coding size of
$A$, so that our transformation is indeed pseudo-polynomial.

Before specifying bound $K(A)$ for the instance of \modul, we will
show three properties of maximum modularity clusterings of $G(A)$.
Together these properties establish the desired characterization of
solutions for \parti\ by solutions for \modul.

\begin{lemma}
In a maximum modularity clustering of $G(A)$,
none of the cliques $H_1,\ldots,H_k$ is split.
\end{lemma}
\begin{proof}
We consider a clustering $\clusteringSymb$ that splits a clique
$H\in\{H_1,\ldots,H_k\}$ into different clusters and then show how to
obtain a clustering with strictly higher modularity. Suppose that
$C_1,\ldots,C_r\in\clusteringSymb$, $r>1$, are the clusters that
contain vertices of $H$. For $i = 1, \ldots, r$ we denote by 
\begin{itemize}
\item $n_i$ the number of vertices of $H$ contained in 
  cluster $C_i$,
\item $m_i=|E(C_i)|$ the number edges between vertices in $C_i$,
\item $f_i$ the number of edges between vertices of $H$ in $C_i$ and
  element vertices in $C_i$,
\item $d_i$ be the sum of degrees of all vertices in $C_i$.
\end{itemize}
The contribution of $C_1,\ldots,C_r$
to $\modularity[\clusteringSymb]$ is
\[ \frac{1}{m} \sum_{i=1}^r m_i - \frac{1}{4m^2} \sum_{i=1}^r d_i^2~. \]
Now suppose we create a clustering $\clusteringSymb'$ by rearranging
the vertices in $C_1,\ldots,C_r$ into clusters $C',C'_1,\ldots,C'_r$,
such that $C'$ contains exactly the vertices of clique $H$, and each
$C'_i$, $1\leq i\leq r$, the remaining elements of $C_i$ (if any). In
this new clustering the number of covered edges reduces by
$\sum_{i=1}^r f_i$, because all vertices from $H$ are removed from the
clusters $C'_i$. This labels the edges connecting the clique vertices
to other non-clique vertices of $C_i$ as inter-cluster edges. For $H$
itself there are $\sum_{i=1}^r\sum_{j=i+1}^r n_in_j$ edges that are
now additionally covered due to the creation of cluster $C'$. In terms
of degrees the new cluster $C'$ contains $a$ vertices of degree
$a$. The sums for the remaining clusters $C'_i$ are reduced by the
degrees of the clique vertices, as these vertices are now in $C'$. So
the contribution of these clusters to $\modularity[\clusteringSymb']$
is given by 
\[ \frac{1}{m} \sum_{i=1}^r \left( m_i + \sum_{j=i+1}^r n_in_j -
  f_i\right) - \frac{1}{4m^2} \left(a^4 + \sum_{i=1}^r 
  (d_i - n_ia)^2\right)~, \] 
so that
\begin{eqnarray*}
\modularity[\clusteringSymb'] - \modularity[\clusteringSymb] & = &
\frac{1}{m} \left(\sum_{i=1}^r \sum_{j=i+1}^r n_in_j -
  f_i\right) + \frac{1}{4m^2} \left(\left(\sum_{i=1}^r 2d_in_ia -
    n_i^2a^2\right) - a^4\right) \\
& = & \frac{1}{4m^2} \left(4m \sum_{i=1}^r \sum_{j=i+1}^r n_i n_j -
  4m\sum_{i=1}^r f_i + \left(\sum_{i=1}^r n_i \left(2d_ia -
      n_ia^2\right)\right) - a^4 \right)
\end{eqnarray*}
Using the fact that $2\sum_{i=1}^r \sum_{j=i+1}^r n_i n_j =
\sum_{i=1}^r \sum_{j \neq i} n_i n_j$, substituting $m =
\frac{k}{2}a(a+1)$ and rearranging terms we get
\begin{eqnarray*}
\modularity[\clusteringSymb'] - \modularity[\clusteringSymb]
& = & \frac{a}{4m^2} \left(- a^3 - 2k(a+1)\sum_{i=1}^r f_i +
  \sum_{i=1}^r n_i \left(2d_i - n_ia + k(a+1)\sum_{j \neq i} n_j
  \right)\right) \\ 
& \ge & \frac{a}{4m^2} \left(- a^3 - 2k(a+1)\sum_{i=1}^r f_i +
  \sum_{i=1}^r n_i \left(n_ia + 2kf_i + k(a+1)\sum_{j \neq i}^r n_j
  \right)\right).
\end{eqnarray*}
For the last inequality we use the fact that $d_i \ge n_ia +
kf_i$. This inequality holds because $C_i$ contains at least the $n_i$
vertices of degree $a$ from the clique $H$. In addition it contains
both the clique and element vertices for each edge counted in $f_i$.
For each such edge there are $k-1$ other edges connecting the element
vertex to the $k-1$ other cliques. Hence, we get a contribution of 
$kf_i$ in the degrees of the element vertices. Combining the terms
$n_i$ and one of the terms $\sum_{j \neq i} n_j$ we get 
\begin{eqnarray*}
\modularity[\clusteringSymb'] - \modularity[\clusteringSymb] & \ge &
\frac{a}{4m^2} \left(- a^3 - 2k(a+1)\sum_{i=1}^r f_i  + \sum_{i=1}^r
  n_i \left(a \sum_{j=1}^r n_j + 2kf_i + ((k-1)a + k)\sum_{j\neq i}^r
    n_j \right)\right) \\ 
& = & \frac{a}{4m^2} \left(- 2k(a+1)\sum_{i=1}^r f_i + \sum_{i=1}^r
  n_i\left(2kf_i + ((k-1)a + k)\sum_{j\neq i}^r n_j \right)\right) \\
& = & \frac{a}{4m^2} \left(\sum_{i=1}^r 2kf_i (n_i - a - 1)) +
  ((k-1)a+k)\sum_{i=1}^r \sum_{j\neq i}^r n_i n_j \right) \\
& \ge & \frac{a}{4m^2} \left(\sum_{i=1}^r 2kn_i (n_i - a - 1) +
  ((k-1)a+k)\sum_{i=1}^r \sum_{j\neq i}^r n_i n_j \right),
\end{eqnarray*}
For the last step we note that $n_i \le a-1$ and $n_i - a - 1 < 0$
for all $i=1,\ldots,r$. So increasing $f_i$ decreases the modularity
difference. For each vertex of $H$ there is at most one edge to a
vertex not in $H$, and thus $f_i \le n_i$.\\
By rearranging and using the fact that $a \ge 3k$ we get
\begin{eqnarray*}
\modularity[\clusteringSymb'] - \modularity[\clusteringSymb] & \ge &
\frac{a}{4m^2} \sum_{i=1}^r n_i \left(2k(n_i - a - 1) +
  ((k-1)a+k)\sum_{j\neq i}^r n_j \right), \\ 
& = & \frac{a}{4m^2} \sum_{i=1}^r n_i \left(- 2k + ((k-1)a-k)\sum_{j\neq
    i}^r n_j \right), \\
& \ge & \frac{a}{4m^2} ((k-1)a-3k) \sum_{i=1}^r \sum_{j\neq i}^r n_i n_j,
\\
& \ge & \frac{3k^2}{4m^2} (3k-6) \sum_{i=1}^r \sum_{j\neq i}^r n_i n_j,
\\
& > & 0,
\end{eqnarray*}
as we can assume $k > 2$ for all relevant instances of \parti. This
shows that any clustering can be improved by merging each clique
completely into a cluster. This proves the lemma.  
\qed\end{proof}

Next, we observe that the optimum clustering places at most one clique
completely into a single cluster. 

\begin{lemma}
In a maximum modularity clustering of $G(A)$, 
every cluster contains at most one of the cliques $H_1,\ldots,H_k$.
\end{lemma}
\begin{proof}
Consider a maximum modularity clustering. The previous lemma shows
that each of the $k$ cliques $H_1,\ldots,H_k$ is entirely contained in
one cluster.  Assume that there is a cluster $C$ which contains at
least two of the cliques. 
%
If $C$ does not contain any element vertices, then the cliques form
disconnected components in the cluster. In this case it is easy to see
that the clustering can be improved by splitting $C$ into distinct
clusters, one for each clique. In this way we keep the number of edges
within clusters the same, however, we reduce the squared degree sums
of clusters.\\
Otherwise, we assume
$C$ contains $l>1$ cliques completely and in addition some element
vertices of elements $a_j$ with $j \in J \subseteq
\{1,\ldots,k\}$. Note that inside the $l$ cliques $\frac{l}{2}a(a-1)$
edges are covered. In addition, for every element vertex corresponding
to an element $a_j$ there are $la_j$ edges included. The degree sum of
the cluster is given by the $la$ clique vertices of degree $a$ and
some number of element vertices of degree $ka_j$. The contribution of
$C$ to $\modularity[\clusteringSymb]$ is thus given by  
\[ \frac{1}{m} \left(\frac{l}{2}a(a-1) + l\sum_{j \in J} a_j\right) -
\frac{1}{4m^2} \left(la^2 + k\sum_{j \in J} a_j\right)^2.\]
Now suppose we create $\clusteringSymb'$ by splitting $C$ into $C'_1$
and $C'_2$ such that $C'_1$ completely contains a single clique
$H$. This leaves the number of edges covered within the cliques the
same, however, all edges from $H$ to the included element vertices
eventually drop out. The degree sum of $C'_1$ is exactly $a^2$, and so
the contribution of $C'_1$ and $C'_2$ to
$\modularity[\clusteringSymb']$ is given by 
\[ \frac{1}{m} \left(\frac{l}{2}a(a-1) + (l-1)\sum_{j \in J}
  a_j\right) - \frac{1}{4m^2} \left(\left((l-1)a^2 + k\sum_{j \in J}
    a_j\right)^2 + a^4\right).\] 
Considering the difference we note that
\begin{eqnarray*}
\modularity[\clusteringSymb'] - \modularity[\clusteringSymb] & = &
- \frac{1}{m} \sum_{j \in J}a_j + \frac{1}{4m^2}\left((2l-1)a^4 +
  2ka^2\sum_{j \in J}a_j - a^4\right) \\ 
& = & \frac{2(l-1)a^4 + 2ka^2\sum_{j \in J}a_j - 4m\sum_{j \in
    J}a_j}{4m^2} \\ 
& = & \frac{2(l-1)a^4 - 2ka\sum_{j \in J}a_j}{4m^2} \\
& \ge & \frac{9k^3}{2m^2}(9k - 1) \\
& > & 0,
\end{eqnarray*}
as $k > 0$ for all instances of \parti. 

%
Since the clustering is improved in each case, it is not optimal.
This is a contradiction.
\qed\end{proof}

The previous two lemmas show that any clustering can be strictly
improved to a clustering that contains $k$ \emph{clique clusters},
such that each one completely contains one of the cliques
$H_1,\ldots,H_k$ (possibly plus some additional element vertices). In
particular, this must hold for the optimum clustering as well. Now
that we know how the cliques are clustered we turn to the element
vertices.\\
As they are not directly connected, it is never optimal to create a
cluster consisting only of element vertices. Splitting such a cluster
into singleton clusters, one for each element vertex, reduces the
squared degree sums but keeps the edge coverage at the same
value. Hence, such a split yields a clustering with strictly higher
modularity. The next lemma shows that we can further strictly improve
the modularity of a clustering with a singleton cluster of an element
vertex by joining it with one of the clique clusters.

\begin{lemma}
In a maximum modularity clustering of $G(A)$, 
there is no cluster composed of element vertices only.
\end{lemma}
\begin{proof}
Consider a clustering $\clusteringSymb$ of maximum modularity and
suppose that there is an element vertex $v_i$ corresponding to the
element $a_i$, which is not part of any clique cluster. As argued
above we can improve such a clustering by creating a singleton cluster
$C = \{v_i\}$. Suppose $C_{min}$ is the clique cluster, for which
the sum of degrees is minimal. We know that $C_{min}$ contains all
vertices from a clique $H$ and eventually some other element vertices
for elements $a_j$ with $j \in J$ for some index set $J$. The cluster
$C_{min}$ covers all $\frac{a(a-1)}{2}$ edges within $H$ and $\sum_{j
  \in J} a_j$ edges to element vertices. The degree sum is $a^2$ for
clique vertices and $k\sum_{j \in J} a_j$ for element vertices. As $C$
is a singleton cluster, it covers no edges and the degree sum is
$ka_i$. This yields a contribution of $C$ and $C_{min}$ to
$\modularity[\clusteringSymb]$ of
\[ \frac{1}{m} \left(\frac{a(a-1)}{2} + \sum_{j \in J} a_j\right) -
\frac{1}{4m^2}\left(\left(a^2 + k\sum_{j\in J} a_j\right)^2 +
  k^2a_i^2\right).\]
Again, we create a different clustering $\clusteringSymb'$ by joining
$C$ and $C_{min}$ to a new cluster $C'$. This increases the edge
coverage by $a_i$. The new cluster $C'$ has the sum of degrees of both
previous clusters. The contribution of $C'$ to
$\modularity[\clusteringSymb']$ is given by
\[ \frac{1}{m} \left(\frac{a(a-1)}{2} + a_i + \sum_{j \in J} a_j\right) -
\frac{1}{4m^2}\left(a^2 + ka_i + k\sum_{j\in J} a_j\right)^2,\]
so that
\begin{eqnarray*}
\modularity[\clusteringSymb'] - \modularity[\clusteringSymb] & = &
\frac{a_i}{m} - \frac{1}{4m^2}\left(2ka^2a_i + 2k^2a_i\sum_{j \in
    J}a_j \right) \\ 
& = & \frac{1}{4m^2}\left(2ka(a+1)a_i - 2ka^2a_i - 2k^2a_i\sum_{j \in
    J}a_j\right) \\ 
& = & \frac{a_i}{4m^2}\left(2ka - 2k^2\sum_{j \in J}a_j\right).
\end{eqnarray*}
At this point recall that $C_{min}$ is the clique cluster with the
minimum degree sum. For this cluster the elements corresponding to
included element vertices can never sum to more than $\frac{1}{k}
a$. In particular, as $v_i$ is not part of any clique cluster, the
elements of vertices in $C_{min}$ can never sum to more than
$\frac{1}{k}(a-a_i)$. Thus, 
\[ \sum_{j \in J} a_j \le \frac{1}{k} (a-a_i) < \frac{1}{k} a, \]
and so $\modularity[\clusteringSymb'] - \modularity[\clusteringSymb] >
0$. This contradicts the assumption that $\clusteringSymb$ is
optimal. 
\qed\end{proof}

We have shown that for the graphs $G(A)$ the clustering of maximum
modularity consists of exactly $k$ clique clusters, and each element
vertex belongs to exactly one of the clique clusters. Finally, we are
now ready to state our main result. 

\begin{theorem}
\modul\ is strongly NP-complete.
\end{theorem}
\begin{proof}
For a given clustering $\clusteringSymb$ of $G(A)$ we can check in
polynomial time whether $\modularity[\clusteringSymb] \ge K(A)$, so
clearly \modul\ $\in$ NP.

For NP-completeness we transform an instance $A=\{a_1,\ldots,a_{3k}\}$
of \parti\ into an instance $(G(A),K(A))$ of \modul. We have already
outlined the construction of the graph $G(A)$ above. For the correct
parameter $K(A)$ we consider a clustering in $G(A)$ with the
properties derived in the previous lemmas, i.e.~a clustering with
exactly $k$ clique clusters. Any such clustering yields exactly
$(k-1)a$ inter-cluster edges, so the edge coverage is given by 
\[ \sum_{C \in \clusteringSymb^*} \frac{|E(C)|}{m} = \frac{m -
  (k-1)a}{m} = 1 - \frac{2(k-1)a}{ka(a+1)} = 1 - \frac{2k-2}{k(a+1)}.
\]
Hence, the clustering $\clusteringSymb = (C_1,\ldots,C_k)$ with
maximum modularity must minimize
\[ \deg(C_1)^2 + \deg(C_2)^2 + \ldots + \deg(C_k)^2.\]
This requires to equilibrate the element vertices according to their
degree as good as possible between the clusters. In the optimum case
we can assign each cluster element vertices corresponding to elements
that sum to $b = \frac{1}{k}a$. In this case the sum of degrees of
element vertices in each clique cluster is equal to $k\frac{1}{k}a =
a$. This yields $\deg(C_i) = a^2+a$ for each clique cluster $C_i$,
$i = 1, \ldots,k$, and gives 
\[\deg(C_1)^2 + \ldots + \deg(C_k)^2 \ge k(a^2 + a)^2 =
ka^2(a+1)^2.\]
Equality holds only in the case, in which an assignment of
$b$ to each cluster is possible. Hence, if there is a clustering
$\clusteringSymb$ with $\modularity[\clusteringSymb]$ of at least
\[
K(A) = 1 - \frac{2k-2}{k(a+1)} - \frac{ka^2(a+1)^2}{k^2a^2(a+1)^2} =
\frac{(k-1)(a-1)}{k(a+1)}
\]
then we know that this clustering must split the element vertices
perfectly to the $k$ clique clusters. As each element vertex is
contained in exactly one cluster, this yields a solution for the 
instance of \parti. With this choice of $K(A)$ the instance
$(G(A),K(A))$ of \modul\ is satisfiable only if the instance $A$ of
\parti\ is satisfiable.\\

Otherwise, suppose the instance for \parti\ is satisfiable. Then
there is a partition into $k$ sets such that the sum over each set
is $\frac{1}{k}a$. If we cluster the corresponding graph by joining
the element vertices of each set with a different clique, we get a
clustering of modularity $K(A)$. This shows that the instance $(G(A),
K(A))$ of \modul\ is satisfiable if the instance $A$ of \parti\ is
satisfiable. This completes the reduction and proves the theorem.
\qed\end{proof}

\section{Conclusion}

We have shown that maximizing the popular modularity clustering index is
strongly NP-complete. These results can be generalized to modularity
in weighted graphs. We can consider the graph $G$ to be completely
connected and use weights of 0 and 1 on each edge to indicate its
presence. Instead of the numbers of edges the definition of modularity
then employs the sum of edge weights for edges within clusters,
between clusters and in the total graph. This yields an equivalent
definition of modularity for graphs, in which the existence of an edge
is modeled with binary weights. An extension of modularity to
arbitrarily weighted graphs is then straightforward. Our hardness
result holds also for the problem of maximizing modularity in weighted
graphs, as this more general problem class includes the problem
considered in this paper as a special case. \\ 
Our hardness result shows that there is no polynomial-time algorithm
optimizing modularity unless P~=~NP. Recently proposed
algorithms~\cite{n-fadcs-03,cnm-fcsln-04,n-mcsn-06,ws-scacg-03,gsa-mfrg-04,rb-smcd-06,da-cdeo-05}
are therefore incorrect in the sense that they yield suboptimal
solutions on many instances. Furthermore, it is a justification to use
approximation algorithms and heuristics to cope with the
problem. Future work includes a deeper formal analysis of the
properties of modularity and the development of algorithms with
performance guarantees.

\subsection*{Acknowledgement}
The authors would like to thank Morten Kloster for pointing out a
mistake in the problem definition of \parti\ in an earlier draft of this
paper.

\bibliographystyle{plain}

\end{document}